\documentclass[conference]{IEEEtran}
\IEEEoverridecommandlockouts
\usepackage{cite}
\usepackage{amsmath,amssymb,amsfonts}
\usepackage{algorithmic}
\usepackage{graphicx}
\usepackage{textcomp}
\usepackage{xcolor}
\usepackage{multicol}
\usepackage{multirow}
\usepackage{subfig}
\usepackage{caption}
\usepackage{tabularx}
\usepackage{hyperref}
\usepackage{makecell}
\usepackage{pdfpages}
    \newcolumntype{L}{>{\raggedright\arraybackslash}X}

\usepackage{dblfloatfix}
\def\BibTeX{{\rm B\kern-.05em{\sc i\kern-.025em b}\kern-.08em
T\kern-.1667em\lower.7ex\hbox{E}\kern-.125emX}}

\hypersetup{
    colorlinks=true,
    urlcolor=blue,
    }

\begin{document}

\title{\huge Reconfigurable Architecture for Spatial Sensing in Wideband Radio Front-End}
\author{M. Gupta, S. Sharma, H. Joshi, and S. J. Darak
 	\thanks{M. Gupta, and S. Sharma are joint first-authors.}
 	\thanks{All authors are with ECE Department, IIIT-Delhi, India 110020, e-mail: \{mansi17165,somya17198, himani,sumit\}@iiitd.ac.in. The work is supported by Core Research Grant (CRG/2019/002568) from DST-SERB, Govt. of India.}
 }

\maketitle
\begin{abstract}
The deployment of cellular spectrum in licensed, shared and unlicensed spectrum demands wideband sensing over non-contiguous sub-6 GHz spectrum. To improve the spectrum and energy efficiency, beamforming and massive multi-antenna systems are being explored which demand spatial sensing i.e. blind identification of vacant frequency bands and direction-of-arrival (DoA) of the occupied bands. We propose a reconfigurable architecture to perform spatial sensing of multi-band spectrum digitized via wideband radio front-end comprising of the sparse antenna array (SAA) and Sub-Nyquist Sampling (SNS). Proposed architecture comprises SAA pre-processing and algorithms to perform spatial sensing directly on SNS samples. The proposed architecture is realized on Zynq System on Chip (SoC), consisting of the ARM processor and FPGA, via hardware-software co-design (HSCD). Using the dynamic partial reconfiguration (DPR), on-the-fly switching between algorithms depending on the number of active signals in the sensed spectrum is enabled. The functionality, resource utilization, and execution time of the proposed architecture are analyzed for various HSCD configurations, word-length, number of digitized samples, signal-to-noise ratio (SNR), and antenna array (sparse/non-sparse).

\end{abstract}


\section{Introduction}
5G is the first cellular network to be deployed in licensed, shared and unlicensed spectrum compared to its predecessors limited to licensed spectrum only. Though 5G is envisioned to operate in sub-6 GHz (below 6 GHz) and mmWave (24 - 52 GHz) spectrum, sub-6 GHz deployment itself demands wideband non-contiguous spectrum sensing. To improve the spectrum and energy efficiency, beamforming and massive multi-antenna systems are being explored which allows spatially separated users to communicate simultaneously over the same frequency. To bring this to reality, a base station with the capability of spatial sensing, i.e. identification of vacant frequency bands and direction-of-arrival (DoA) of the occupied bands \cite{doa_wireless},  over the non-contiguous spectrum are desired. Spatial sensing is also useful in radar, sonar, and navigation to track an object \cite{doa_radar}, biomedical to detect tumors, and artery wall movement \cite{doa_biomedical}. Emerging applications demand low execution time i.e. fast sensing which demands acceleration on dedicated hardware such as co-processors, or Field Programmable Gate Arrays (FPGAs). In a dynamic environment, a single algorithm may not always offer superior performance and hence, reconfigurable architectures are being explored.



Various works have discussed the efficient hardware-based spatial sensing \cite{DoA_hw1,DoA_hw2,b4,b5,b6,b7}. Among them, \cite{b4,b5,b6,b7} consider the acceleration by implementing the spatial sensing on FPGA. In \cite{b4}, two LU decomposition-based methods for Uniform Linear Array (ULA) are presented along with performance analysis with the QR decomposition method. In \cite{b5}, the FPGA implementation of the Multiple Signal Classification (MUSIC) algorithm is presented. Other works include DOA estimation using  Estimation of Signal Parameters via Rotational Invariance Techniques (ESPRITs) algorithm \cite{b6} and Barlett algorithm\cite{b7}. Existing works may not be suitable for next-generation networks due to the high sampling rate for digitization of the wideband spectrum and hence, non-contiguous Sub-Nyquist sampling (SNS) is preferred \cite{anil1, cascade, isj}. Furthermore, area and power requirements limit the number of physical antennas and hence, sparse antenna arrays (SAA) are being explored \cite{b1}. Conventional spatial sensing cannot work directly with SNS and SAA-based wideband radio front-end (WRFE) due to the requirement to compensate for the loss of samples and fewer physical antennas. Furthermore, reconstruction of the Nyquist ULA spectrum may not be efficient, and hence, how to minimize the overhead due to SAA and SNS is an important research direction \cite{anil1}.  


In this paper, we propose a reconfigurable architecture for spatial sensing of multi-band spectrum digitized via WRFE. Proposed architecture includes SAA pre-processing (SAP) to compensate for fewer physical antennas and algorithms which perform spatial sensing directly on SNS samples. The proposed architecture is realized on Zynq System on Chip (SoC), consisting of the ARM processor, NEON co-processor, and FPGA, via hardware-software co-design (HSCD). Using the dynamic partial reconfiguration (DPR), on-the-fly switching between algorithms depending on the number of active signals is accomplished. The functionality, execution time, resource, and power consumption are analyzed for various HSCD configurations, word-length, number of samples,  signal-to-noise ratio (SNR), and antenna array. Please refer to \cite{SC} for source codes and a detailed tutorial. Next, we present the WRFE model. 

\section{Wideband Radio Front-End Model}
\label{Sec:System_Model}
\begin{figure}[!t]
        \centering                   
        \includegraphics[scale=0.8]{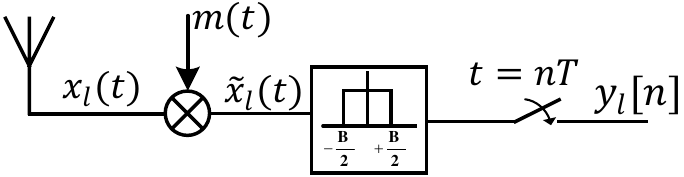}
        \caption{\footnotesize $l^{th}$ antenna receiver of the SNS Architecture {\cite{cascade}.}}
        \label{fig:Rx}
        \vspace{-0.5cm}
\end{figure}

Consider a wideband spectrum consisting of multiple disjoint, uncorrelated and far-field signals \cite{b1,anil1,cascade}. These signals impinge on the $L$-antenna receiver with either ULA or SAA. The signal received at the $l^{th}$ antenna is given as
\begin{align}
    x_l(t) & = \sum_{m=1}^M a_m(t-\tau_l(\theta_m)) e^{j2\pi f_m (t-\tau_l(\theta_m))} + n_l(t)\\
        & \approx \sum_{m=1}^M a_m(t) e^{j2\pi f_m (t-\tau_l(\theta_m))} + n_l(t)
\end{align}
where $M$ is the number of narrowband signals, $a_m(t)$ is the amplitude of $m^{th}$ narrowband signal of carrier frequency, $f_m$ and DoA, $\theta_m$, $\tau_l(\theta_m)$ is the time delay observed by the $m^{th}$ signal at the $l^{th}$ antenna and $n_l(t)$ is the additive Gaussian noise at the $l^{th}$ antenna. As shown in Fig.~\ref{fig:Rx}, the output of the $l^{th}$ antenna is digitized via SNS where the received signal, $x_l(t)$ is mixed with a mixing function, $m(t) = \sum_{b\in \beta}\alpha_{l,b}e^{-j2\pi (b-1)Bt}$. Here, $\beta \in \{1,2,\cdots,N\}$ is a set of non-contiguous frequency bands over which digitization is performed, $N$ is the total number of non-overlapping frequency bands in the wideband spectrum, $\alpha_{l,b}$ is a mixing coefficient of $b^{th}$ frequency band at $l^{th}$ antenna  selected from the Gaussian distribution, and $B$ is the bandwidth of a frequency band. The Fourier transform (FT) of the mixed signal, $\tilde{x}_l(t)$, is given as \cite{isj}
\begin{equation}
    \tilde{X}_l(f)  
    = \sum_{m=1}^{M} e^{j2\pi f_m\tau_l(\theta_m)}\sum_{b\in \beta}\alpha_{l,b} A_m(f-(f_m-(b-1)B)) 
\end{equation}
where $A_m(f)$ is the FT of $m^{th}$ signal. Since $\tilde{X}_l(f)$ contains images over the entire frequency, the mixed signal is passed through the low pass filter (LPF) of cut-off frequency $B/2$. The filtered signal is digitized at a rate of $B~Hz$. The output of the ADC can be written as
\begin{equation}
	Y_{l}(e^{j2\pi f NT}) = \sum_{b\in\beta_{busy}} \alpha_{l,b} A_{b}(f-(f_{b}-(b-1)B))e^{j2\pi f_{b}\tau_l(\theta_{b})}  
\end{equation}
where $\beta_{busy}\in \beta$ is a selected set of occupied frequency bands, $A_b(f)$ is the FT of the signal in $b^{th}$ frequency band with frequency and DoA of $f_b~Hz$ and $\theta_b$, respectively. The FT of the output of all $L$ antennas in the matrix form is given as 
\begin{equation}
    \textbf{Y}(f) = \textbf{S}~\textbf{Z}(f)      
    \label{freq_domain}
\end{equation}
where $\textbf{S}$ is a $L\times M$ steering matrix where $\textbf{S}_{l,m}=e^{j2\pi f_m \tau_l(\theta_m)}$.
Note that the steering matrix, $\textbf{S}$ contains two variables, carrier frequency, $f_m$ and DoA, $\theta_m$ for each of the narrowband signal. Similar to \cite{b1,DoA_hw1,DoA_hw2,b4,b5,b6,b7}, we focus on DoA estimation with known carrier frequency. 

\section{Proposed Spatial Sensing for WRFE}
\label{Sec:spatial_sensing_wrfe}
\begin{figure}[!t]
        \centering   
        \vspace{-0.2cm}
        \subfloat[]{\includegraphics[scale=0.32]{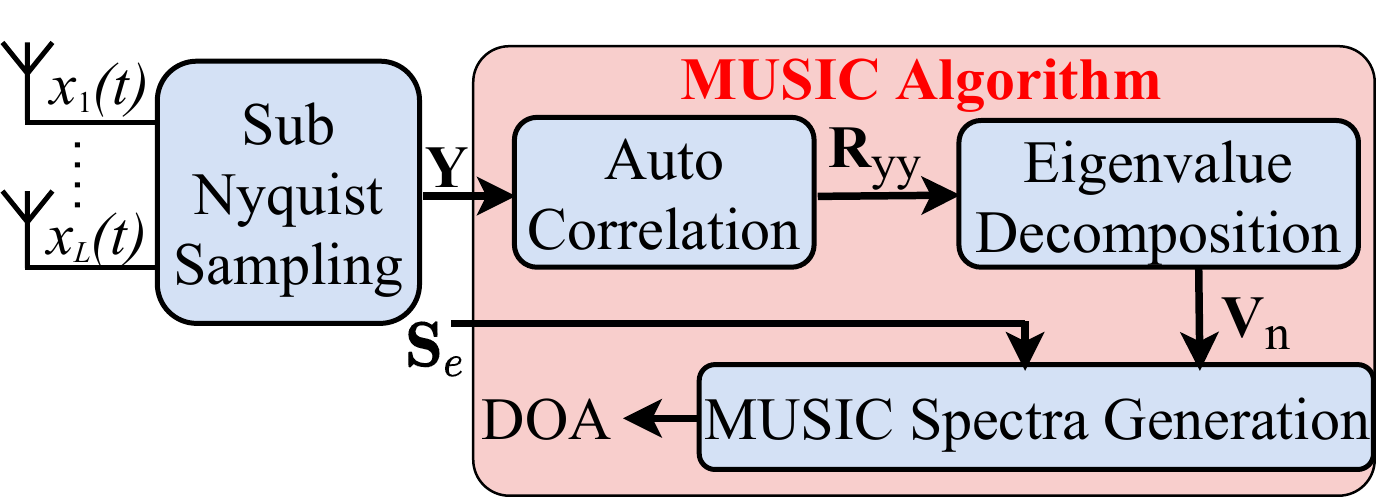}
         \label{fig:ULA_block_diagram}}
        \subfloat[]{\includegraphics[scale=0.3]{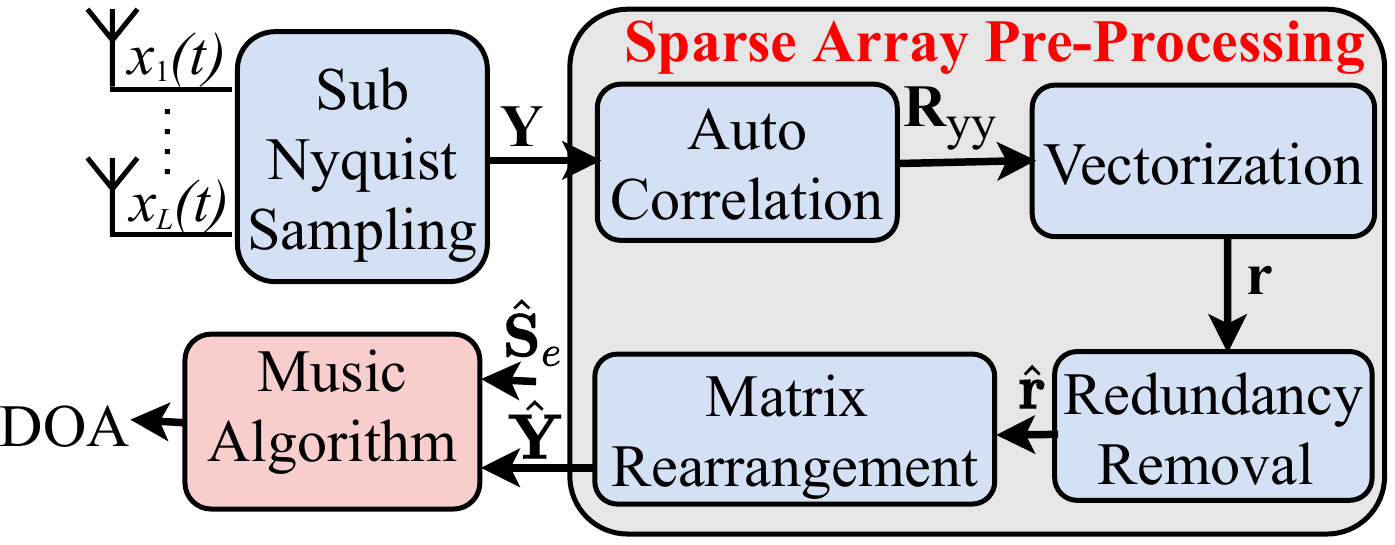}\label{fig:Sparse_block_diagram}}
         \caption{\footnotesize DOA Estimation for (a) ULA and (b) SAA via MUSIC Algorithm.}
           \vspace{-0.4cm}
\end{figure}

The proposed baseband spatial sensing approach to estimate DoA for ULA and SAA based WRFE is shown in Fig.~\ref{fig:ULA_block_diagram} and \ref{fig:Sparse_block_diagram}, respectively.  The inputs are two complex matrices: 1) Sub-Nyquist samples, $\textbf{Y}$ of size $L\times K$ where $K$ denotes the number of baseband samples at the output of SNS and 2) Since  $\textbf{S}$ is unknown, we use extended steering matrix,  $\textbf{S}_{e}$ where $({\textbf{S}}_e)_{l,p}=e^{j2\pi f \tau_l(\theta_p)}$ and  $\theta_p\in[0^\circ,180^\circ]$. 

\subsection{Spatial Sensing for WRFE with ULA}
In ULA, the antennas are uniformly placed, and hence, the time delay at the $l^{th}$  antenna of the steering matrix is given as
\begin{equation}
    \tau_l(\theta_m)  = (l-1)\frac{d}{c}cos(\theta_m)
\end{equation}
where $d$ is the distance between two adjacent antennas and $c$ is the speed of light.
To estimate unknown DoA, $\theta$, an extended steering matrix $\textbf{S}_{e}$ and sub-Nyquist samples, $\textbf{Y}$  are processed via MUSIC algorithm. The output is the MUSIC spectra at every possible $\theta_i$ where $\theta_i\in[0^\circ,180^\circ]$ and it is given as
\begin{equation}
    \label{pmusic_eq}
    P[\theta_i] = \frac{1}{|(\textbf{S}_{e}[:,i]^H\textbf{V}_n)(\textbf{V}_n^H\textbf{S}_{e}[:,i])|}
\end{equation}
where $\textbf{V}_n$ is the eigenvectors of noise subspace of the auto-correlation matrix, $\textbf{R}_{yy}$ of the sub-Nyquist samples, $\textbf{Y}$. To find $\textbf{V}_n$, first step is to calculate auto-correlation matrix $\textbf{R}_{yy}$ followed by Eigenvalue decomposition (EVD). In the end, the DoA of a signal is estimated based on the locations of the peaks identified in the MUSIC spectrum. 

In SAA, the number of physical antennas is less than the number of antenna slots, and such sparsity in antenna placement is exploited to estimate a higher number of DoAs than ULA for a given number of physical antennas. To enable this, a sparse array pre-processing (SAP) is needed which performs rank enhancement of the sub-Nyquist samples matrix, $\textbf{Y}$.
Let $\mathcal{L} \in \{1, 2, \cdots, L^{'}\}$ be a set storing the locations of the antennas of SAA where $L^{'} >L$ is the location of the last antenna and is equal to the total number of antennas i.e. the sum of $L$ physical antennas and virtual antennas.
The input of the SAP block is the sub-Nyquist samples, $\textbf{Y}$ which is passed through auto-correlation function (ACF) to obtain $\textbf{R}_{yy}$ of size $L\times L$. The $1\times L^2$ vector $\textbf{r}$ is obtained by applying Khatri-Rao property of vectorization on $\textbf{R}_{yy}$. In redundancy removal block, the reduced vector $\hat{\textbf{r}}$ of size $1\times (2L^{'}-1)$ is generated by removing the redundant entries of vector $\textbf{r}$. The $L^{'} \times L^{'}$ SAP matrix, $\hat{\textbf{Y}}$, is a spatially smoothed matrix and its $i^{th}$ column is defined as $[\hat{r}_{L^{'}+1-i}, \hat{r}_{L^{'}+2-i}, \cdots, \hat{r}_{L^{'}+L^{'}-i}]^{'}$ where $i = \{1, 2, \cdots, L^{'}\}$ and $\hat{r}_l$ is the $l^{th}$ entry of the reduced vector $\hat{\textbf{r}}$. 
Finally, the SAP matrix, $\hat{\textbf{Y}}$ and the extended steering matrix, $\hat{\textbf{S}}_{e}$ are passed to the MUSIC algorithm to estimate DoAs. Here, the $\{l,p\}^{th}$ entry of the extended steering matrix, $\hat{\textbf{S}}_{e}$, defined as $e^{j2\pi f \tau_l(\theta_p)}$ where $l\in\mathcal{L}$ (for ULA,   $l\in\{1,2,\cdots,L\}$) and  $\theta_p\in[0^\circ,180^\circ]$. 

\section{Reconfigurable Architecture}
\label{Sec:hw_implement}
\begin{figure}[h]
        \centering                   \vspace{-0.1cm}
        \includegraphics[scale=0.575]{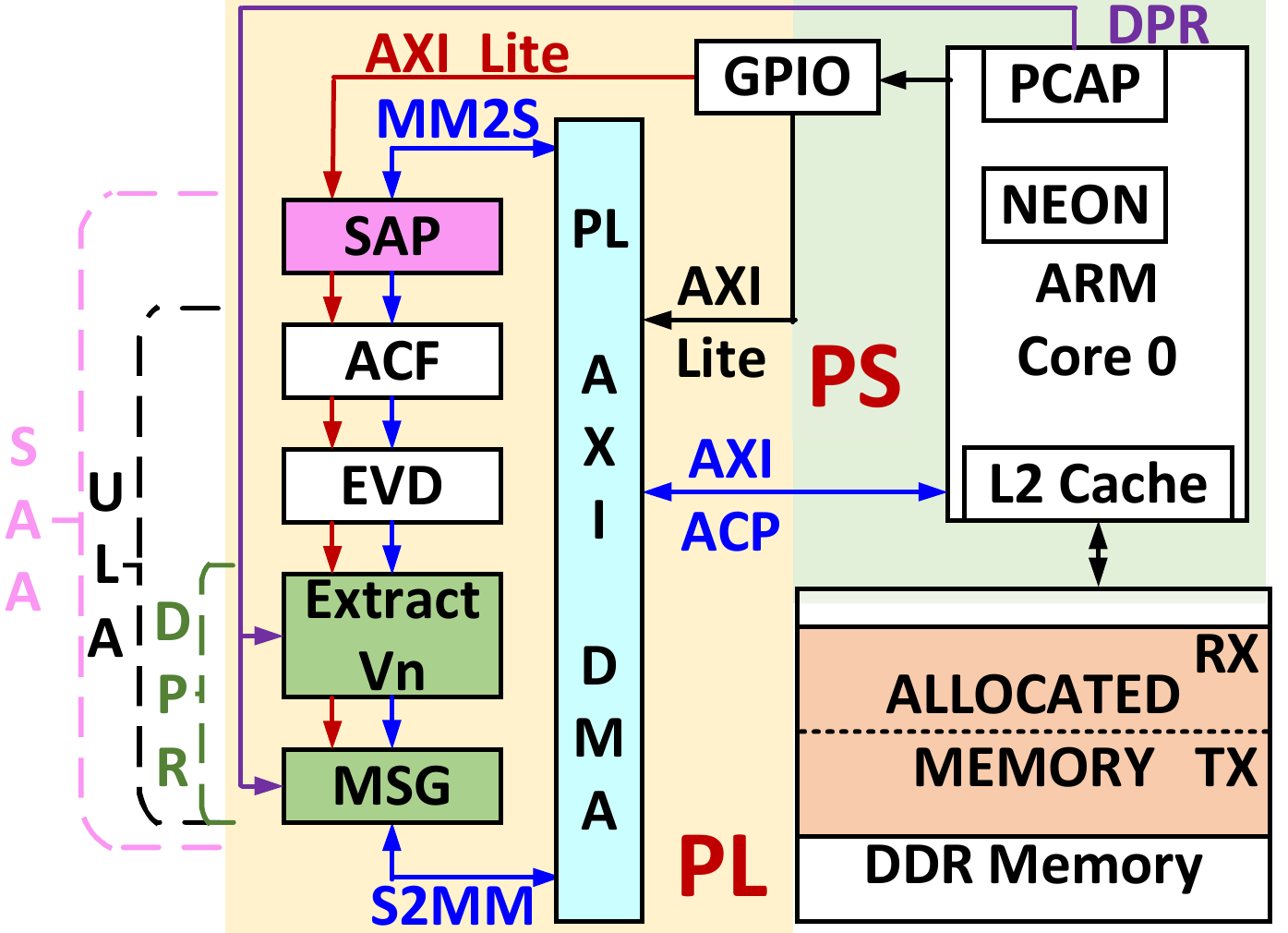}
        \vspace{-0.1cm}
        \caption{Reconfigurable architecture for spatial sensing.}
        \vspace{-0.2cm}
        \label{fig:PS_PL_ARCH}
\end{figure}

The proposed reconfigurable architecture for SAA-based spatial sensing on ZSoC is shown in Fig.~\ref{fig:PS_PL_ARCH}. The data received from WRFE is stored by the ARM Core 0 of the PS in the DDR memory. The PS configures the AXI Direct Memory Access (DMA) in PL so that it reads the data from DDR memory via memory-mapped AXI Accelerator Coherency Port (ACP), forwards it to spatial sensing block via AXI stream interface, and stores the processed data back to the DDR memory. The PS displays the calculated DoA using the UART terminal. The ULA spatial sensing architecture is obtained by removing the SAP block in Fig.~\ref{fig:PS_PL_ARCH}. The architecture is made reconfigurable in $M$ i.e. number of active transmissions. This is done using DPR based on-the-fly configuration of Extract $\textbf{V}_n$ and MSG blocks via processor configuration access port (PCAP). Please refer to \cite{SC} for source codes and a detailed tutorial. In Fig.~\ref{fig:PS_PL_ARCH}, complete spatial sensing is mapped on the PL. Various configurations of PS-PL division via HSCD are also explored and corresponding performance analysis is discussed in Section~\ref{Sec:results}.


\vspace{-0.1cm}
\subsection{Sparse Array Pre-Processing (SAP)}
As discussed in Section~\ref{Sec:spatial_sensing_wrfe} and Fig~\ref{fig:Sparse_block_diagram}, the first step in SAP is the auto-correlation function (ACF) on input matrix $\textbf{Y}$. It is same as the first step in MUSIC as shown in Fig.~\ref{fig:ULA_block_diagram} and its architecture is discussed later in Section~\ref{Sec:ACF}. Next step in SAP is the vectorization which involves reading $\textbf{R}_{yy}$ in column-wise fashion and storing it in vector $\textbf{r}$. This is accomplished using dual-port block RAM in FPGA. In redundancy removal and matrix rearrangement, the redundant entries in $\textbf{r}$ are removed, and the rest are rearranged to form new matrix $\hat{\textbf{Y}}$. The hardware implementation involves load operation from $\textbf{r}$ and store operation at the specific address of $\hat{\textbf{Y}}$ as discussed in Section III.
\vspace{-0.1cm}
\subsection{Auto-Correlation Function (ACF)} 
\label{Sec:ACF}
\begin{figure}[h]
        \centering    
        \vspace{-0.1cm}
        \includegraphics[scale=0.4]{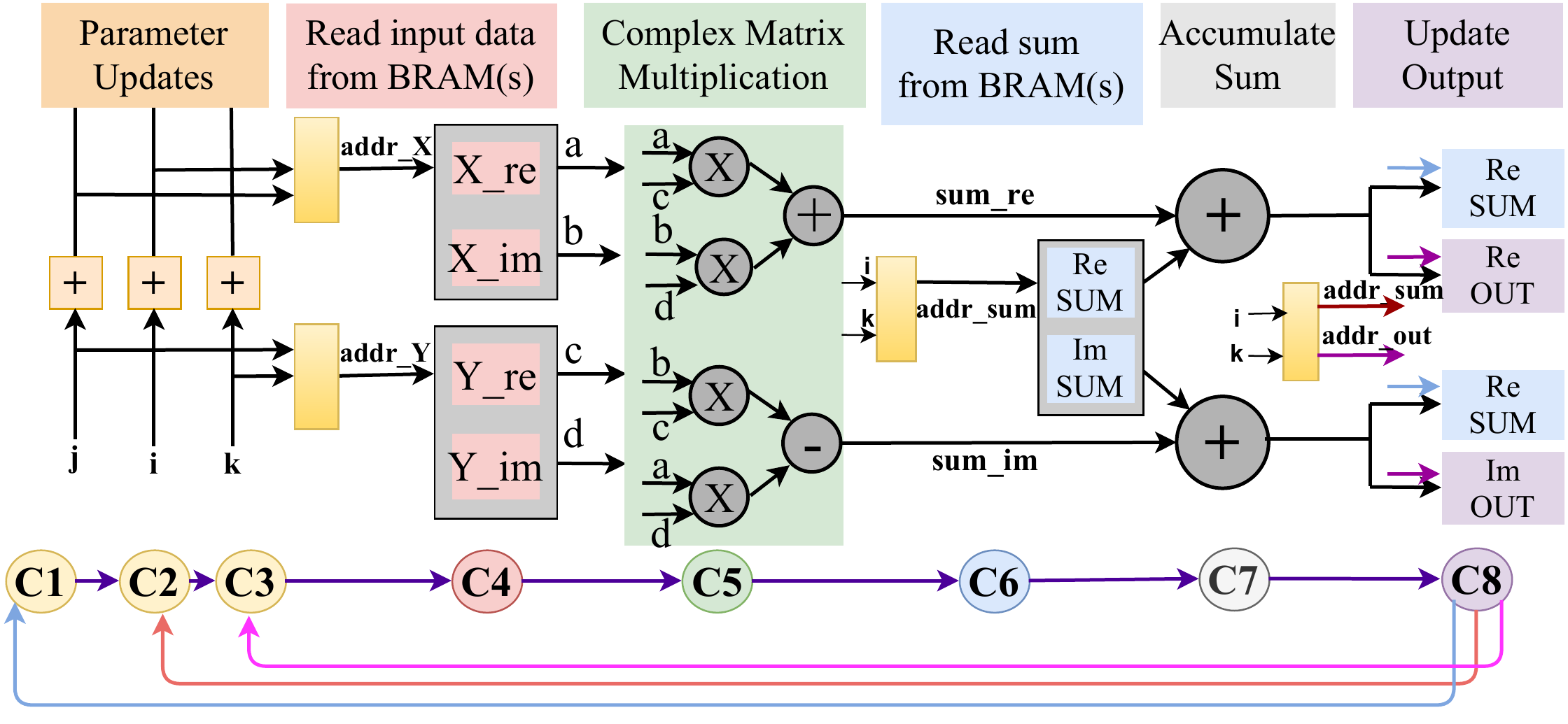}
        \vspace{-0.1cm}
        \caption{Complete ACF Architecture.}
        \vspace{-0.1cm}
        \label{fig:Correlation_ARCH}
\end{figure}

This is the first step in SAP as well as the MUSIC algorithm and involves multiplication of input matrix with its Hermitian. For instance, ACF of matrix $\textbf{Y}$ of size $L\times K$, is obtained as 
\begin{equation}
    \label{AC_eq1}
    \textbf{R}_{yy}=\textbf{Y}*\textbf{Y}^H
\end{equation}
 The size of $\textbf{R}_{yy}$ is $L \times L$. The accumulation of the scalar product of ${(i,j)}^{th}$ element from $\textbf{Y}$ and $j^{th}$ row of $\textbf{Y}^{H}$ for all values of $j$ gives the $i^{th}$ row of $\textbf{R}_{yy}$ matrix. Initially, the matrix $\textbf{Y}$ is copied into the local dual-port BRAM units and appropriate partitioning is done to allow reading and writing of multiple elements simultaneously. As shown in Fig.~\ref{fig:Correlation_ARCH}, FSM($C0$ to $C8$) controls complete ACF execution. The parameters are initialized in the $C0$ state.  In $C1$, $C2$, $C3$ states, the parameters that control the read and write addresses from memory are updated. From $C3$, it enters $C4$ where two complex numbers (one each from $\textbf{Y}$ and $\textbf{Y}^{H}$) are read from the address generated by FSM. This is followed by complex multiplication involving four real multiplications. $C6$ loads the accumulator value from BRAM ($SUM$) and in $C7$, the result from $C5$ is added to the accumulator value. The updated accumulator value is written to either $SUM$ or $OUT$ based on the control signals. 
 From $C8$, FSM switches back to $C3$ $L$ times before switching to $C2$. At this time, a single row of the $SUM$ is updated. Thus, when FSM moves from $C3$ to $C1$, the entire $SUM$ matrix has been initialized. At the end of $K^{th}$ switch from $C8$ to $C1$, all elements of $\textbf{R}_{yy}$ are available. 
 


\vspace{-0.1cm}
\subsection{Eigen Value Decomposition (EVD)} 
The EVD is needed to obtain matrix $V$ by decomposing the auto-correlation matrix, $\textbf{R}_{yy}$ into the signal and noise subspace. 
This is achieved via iterative $QR$-decomposition method to obtain eigenvalues ($\textbf{U}$) and eigenvectors ($\textbf{V}$) of the input auto-correlation matrix.
 The EVD is realized on PL using the QR IP provided by Xilinx, page $275$ of \cite{QR_IP}.
 \vspace{-0.1cm}
\subsection{Reconfigurable $V_n$ Extraction}
Depending on the number of active transmissions $M$, smallest $L-M$ Eigenvalues are identified and corresponding 
Eigenvectors are extracted to form matrix, $\textbf{V}_n$ of size $L\times (L-M)$. Depending on $M$, this block gets configured with appropriate bit-stream via DPR. 
\vspace{-0.1cm}
\subsection{Reconfigurable MUSIC Spectra Generation (MSG)} 
\begin{figure}[!b]
\vspace{-0.2cm}
        \centering                   
        \includegraphics[scale=0.52]{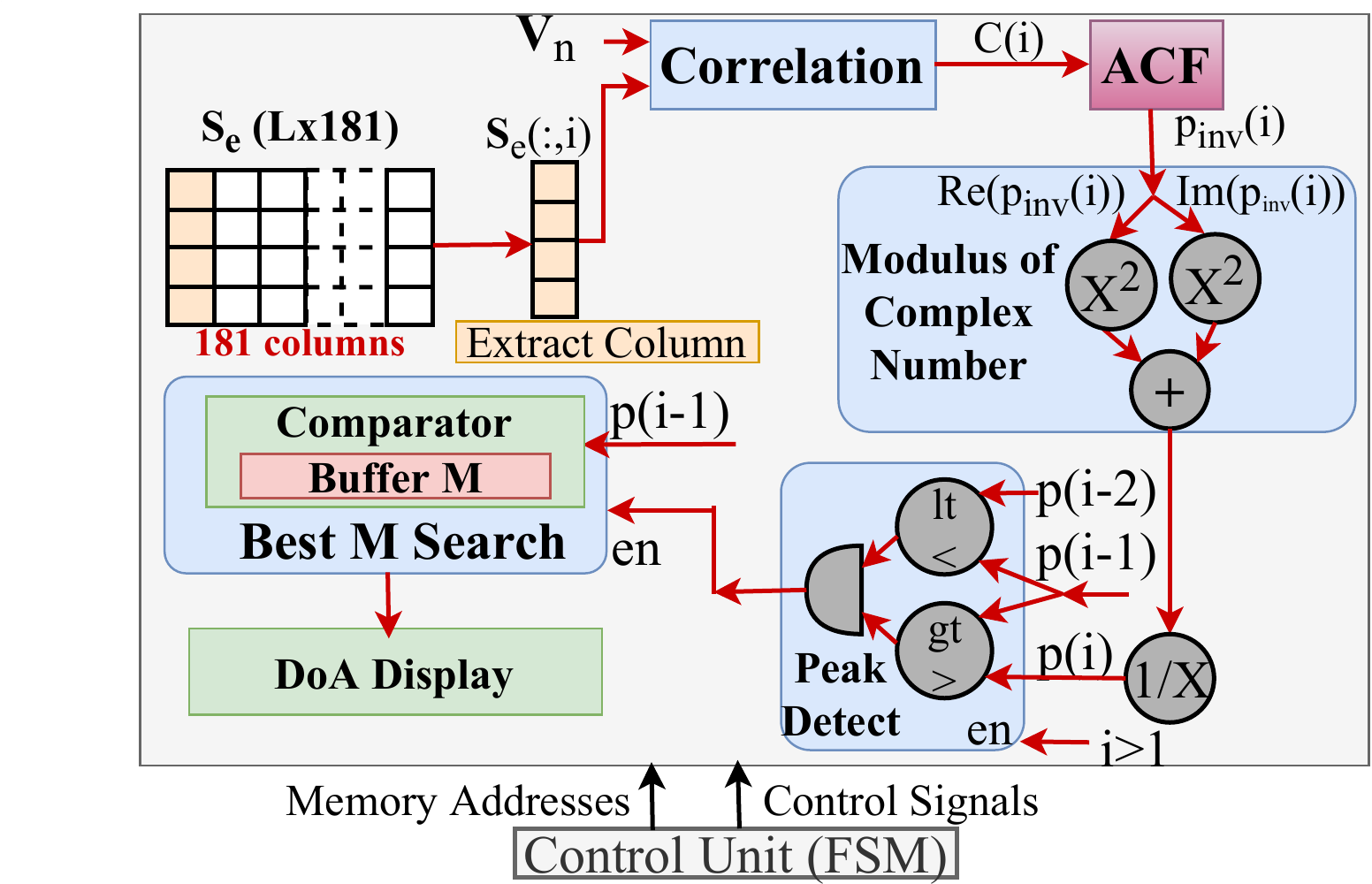}
        \vspace{-0.1cm}
        \caption{Music Spectra Generation Architecture}
        \label{fig:MSG_ARCH}
        \vspace{-0.2cm}
\end{figure}

\begin{figure*}[!b]
        \centering        
        \vspace{-0.2cm}
        \subfloat[]{\includegraphics[scale=0.29]{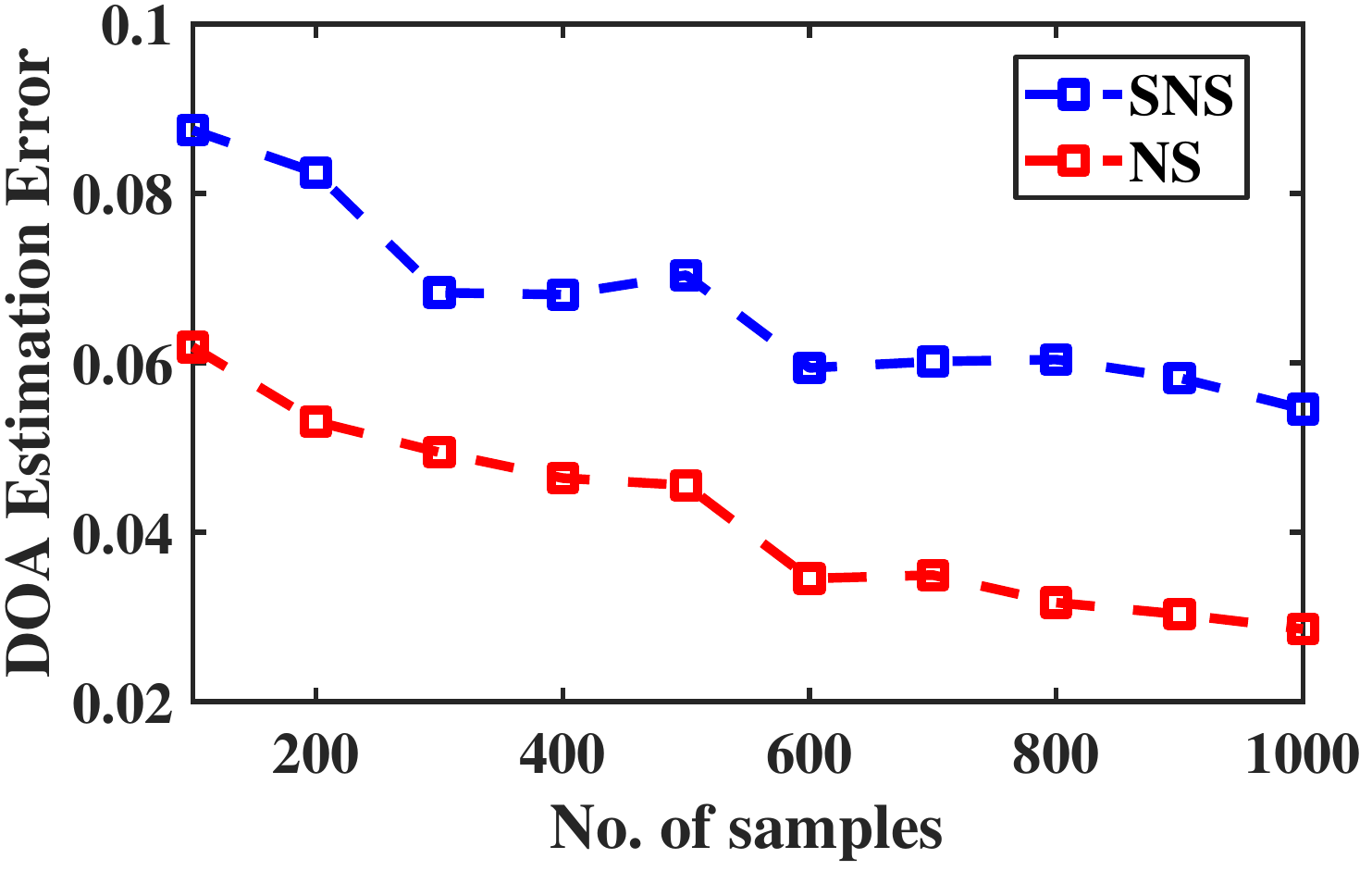}}
        \hspace{.03cm}
        \subfloat[]{\includegraphics[scale=0.28]{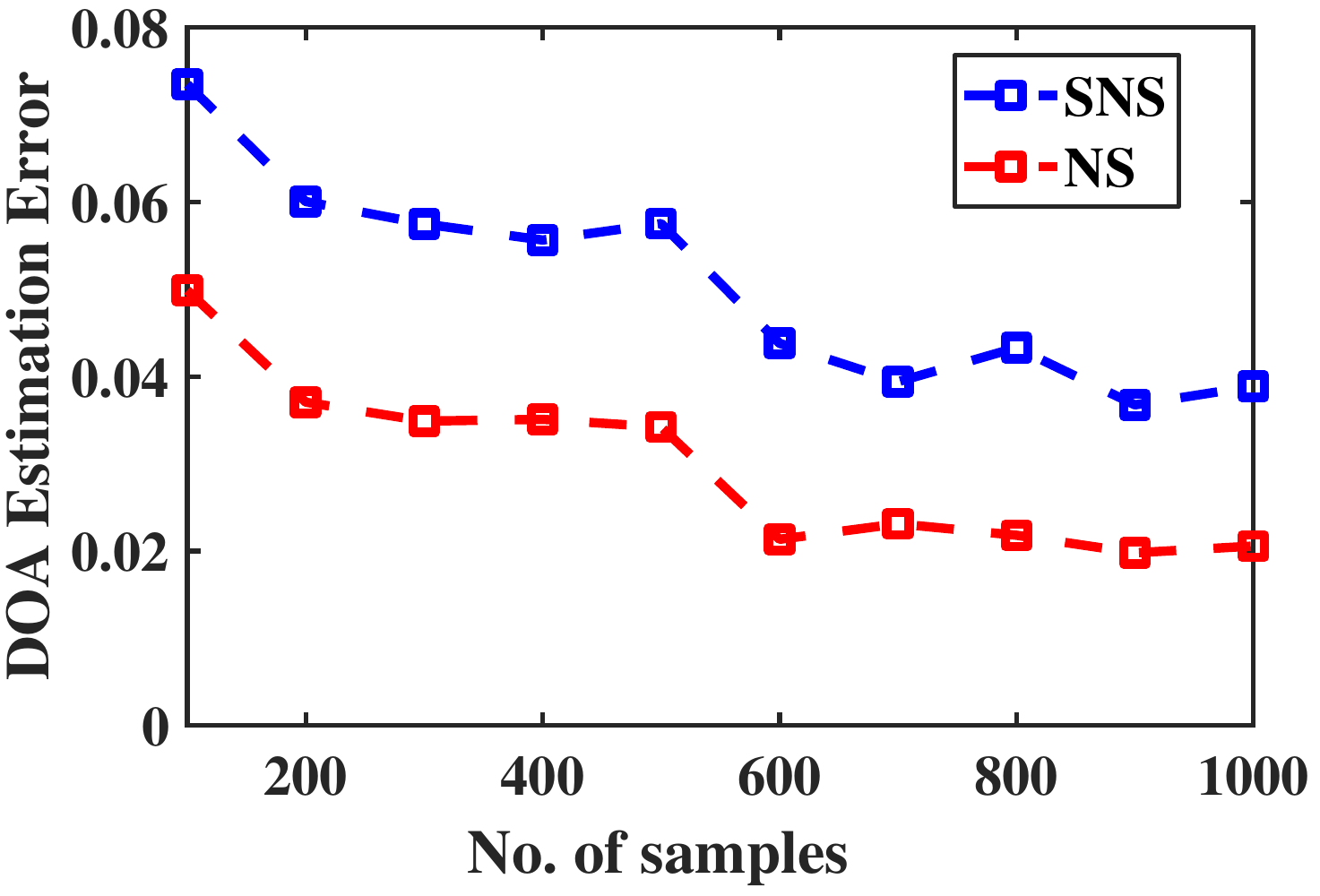}}
        \hspace{.03cm}
         \subfloat[]{\includegraphics[scale=0.29]{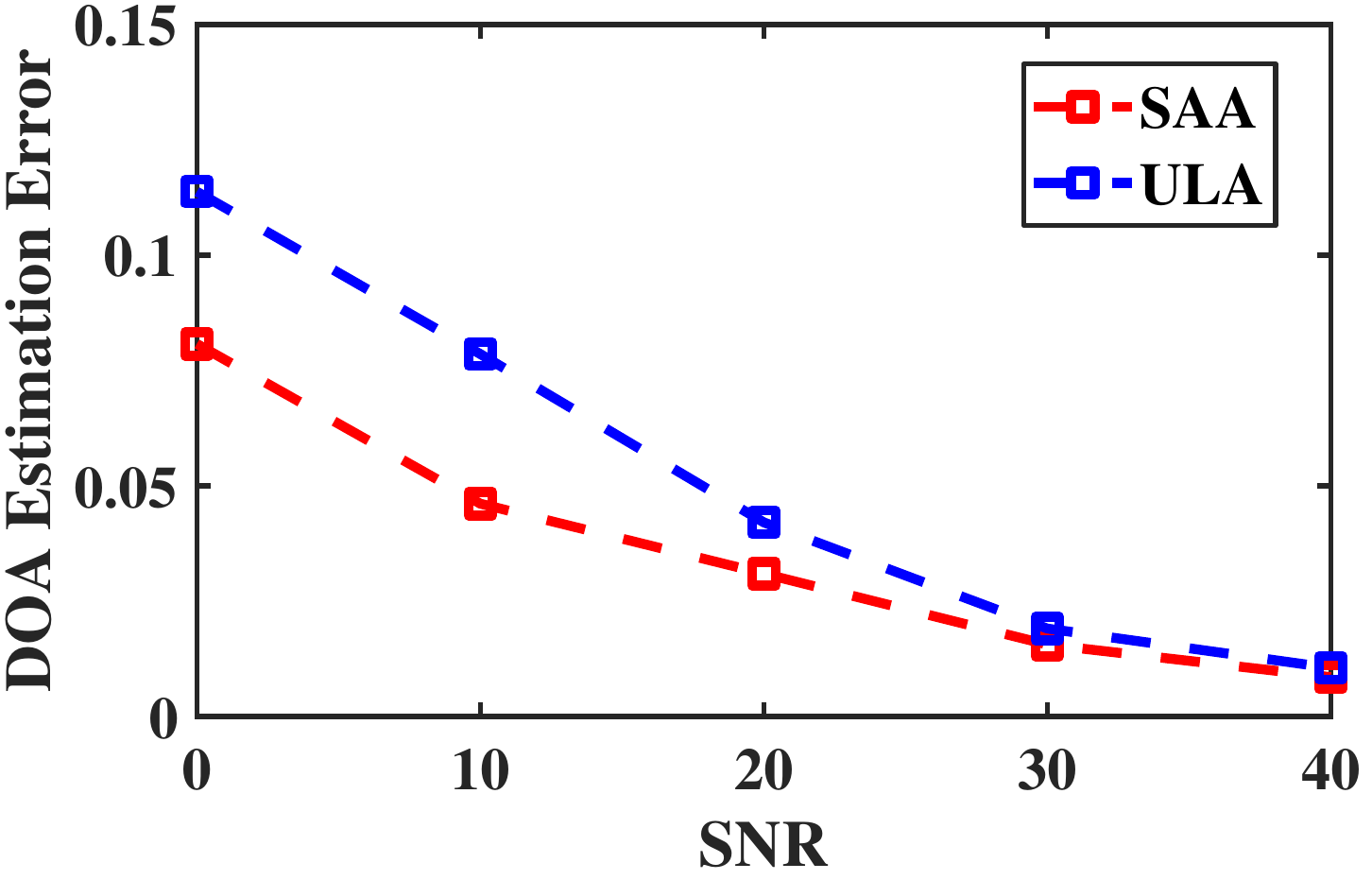}}
         \vspace{-0.1cm}
        \caption{\footnotesize NDEE vs (a) No. of RF Samples for ULA, (b) No. of RF Samples for SAA, and (c)  (b) SNR for ULA and SAA.} 
        \label{fig:ULA_error_sample}
        \vspace{-0.1cm}
\end{figure*}
In MSG block, MUSIC spectrum is generated using the extended steering matrix, $\textbf{S}_{e}$ and the noise eigenvector, $\textbf{V}_n$ as discussed in Eq.~\ref{pmusic_eq}. The indices of the highest $M$ peaks in the MUSIC spectra are the estimated DoA values.



The proposed architecture for MSG is shown in Fig.~\ref{fig:MSG_ARCH}. The $\textbf{S}_{e}$ is partitioned to sequentially extract each column i.e. $\textbf{S}_{e}[:,i]$ where $i\in [0,180]$. The output of the correlation of $\textbf{S}_{e}[:,i]$ and $\textbf{V}_{n}$ is a vector of size $1\times (L-M)$. As shown in Eq.~\ref{pmusic_eq}, next step is the auto-correlation to get single complex value $p_{inv}(i)$ followed by modulus operation to obtain $P(\theta_i)$. This is achieved 
via two multiplication units and one addition unit. Note that the square root operation is skipped as it does not impact the DOA estimation accuracy. 

Next step is to find DoAs present in the input signal and it corresponds to indices of $M$ maximum values of $P(\theta_i)$ where $i\in [0,180]$. The first step is to find out the indices, $i$, which corresponds to the peak. This is achieved by comparing each  $P(\theta_i)$ with its neighbouring values, $P(\theta_{i-1})$ and $P(\theta_{i+1})$ as shown in Fig.\ref{fig:MSG_ARCH}.  In the end, the remaining values are sorted and indices of elements of $P$ having best $M$ values corresponds to DoA in the input signal.

\section{Hardware Results}
\label{Sec:results}


 We begin with the validation of the functional correctness of the proposed architecture by comparing the normalized DOA estimation error (NDEE) with Nyquist sampling-based spatial sensing for ULA and SAA. We consider the effect of the number of RF samples for $L=4$ antennas and SNR of $20~dB$ in Fig.~\ref{fig:ULA_error_sample} (a) and (b) for ULA and SAA, respectively. Note that these RF samples are passed through SNS and NS-based digitization. In SNS, the number of digitized samples, $K$, is 5 times lower than NS. In Fig.~\ref{fig:ULA_error_sample} (c),
 we consider the effect of SNR on NDEE for ULA and SAA.
 As expected, NDEE decreases with the increase in the number of RF samples and SNR. The NDEE of the SNS is slightly higher due to the penalty of digitization via low-rate ADCs and lower $K$ i.e. 5 times fewer baseband samples. However, the NDEE is less than 0.03 i.e. at most 5.4 degree.

\par

\begin{figure}[!t]
        \centering          
        \vspace{-0.3cm}
        \subfloat[]{\includegraphics[scale=0.3]{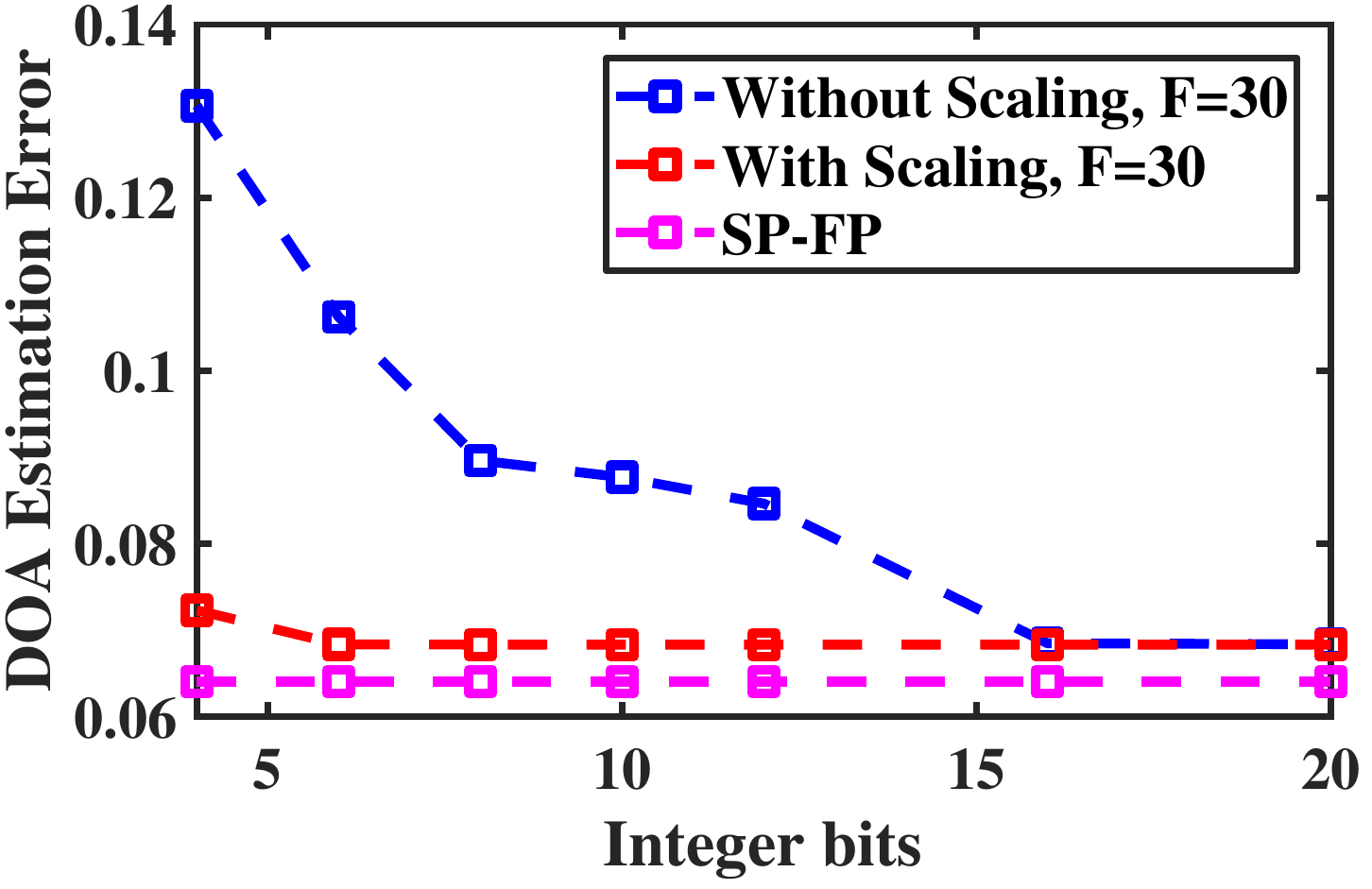}}
        \subfloat[]{\includegraphics[scale=0.3]{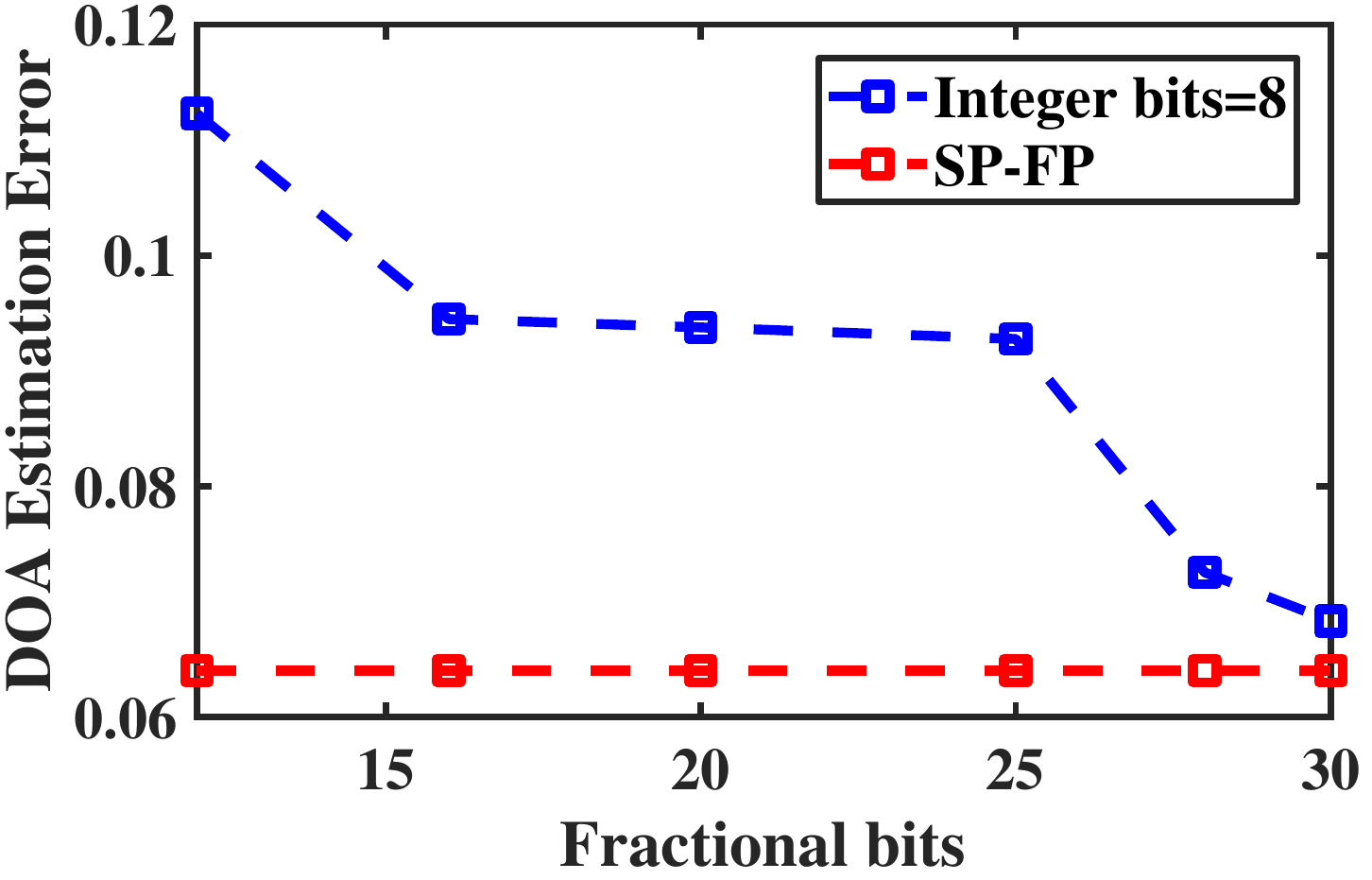}}\\
        \vspace{-0.25cm}
        \subfloat[]{\includegraphics[scale=0.3]{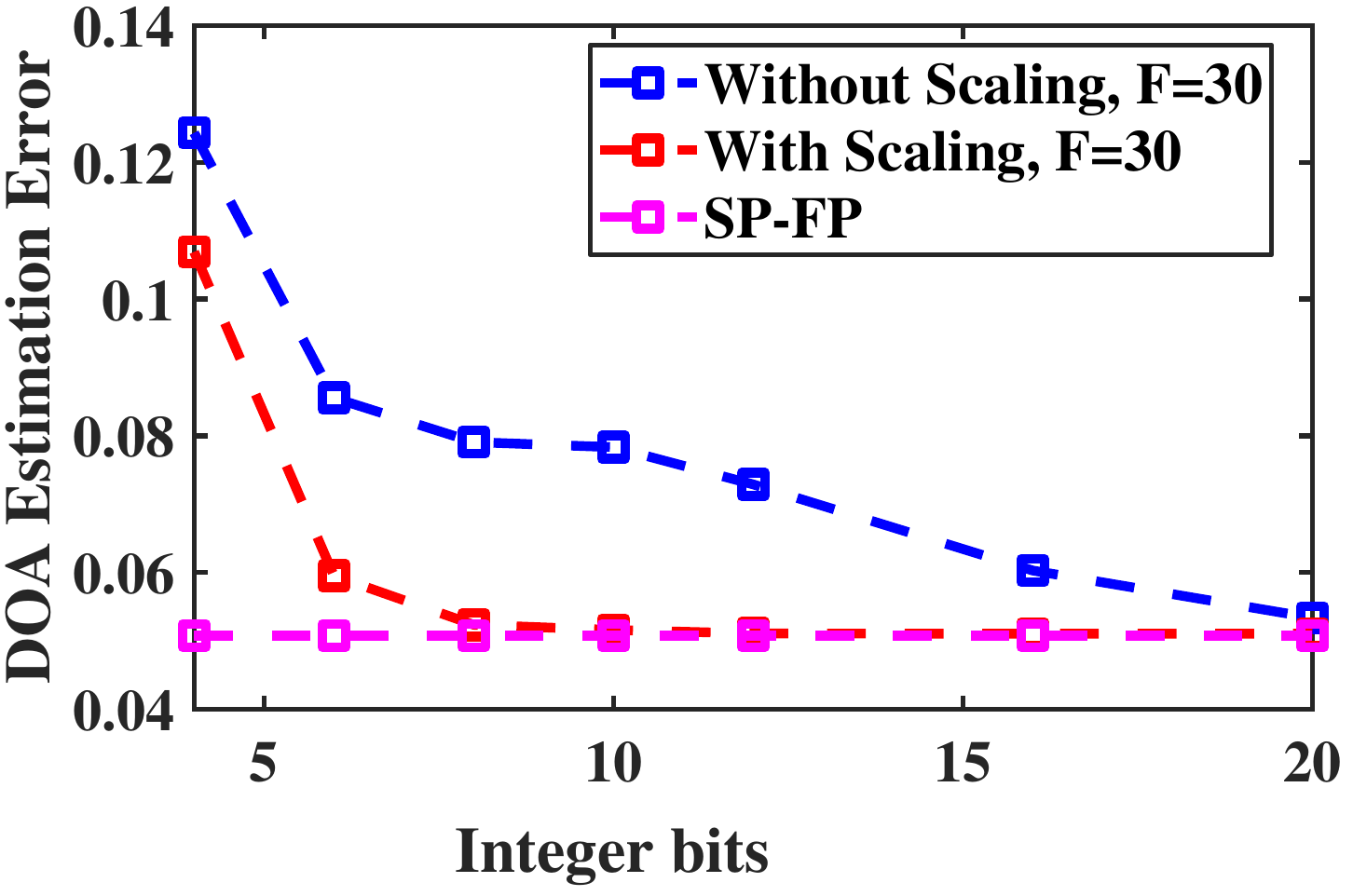}}
        \subfloat[]{\includegraphics[scale=0.3]{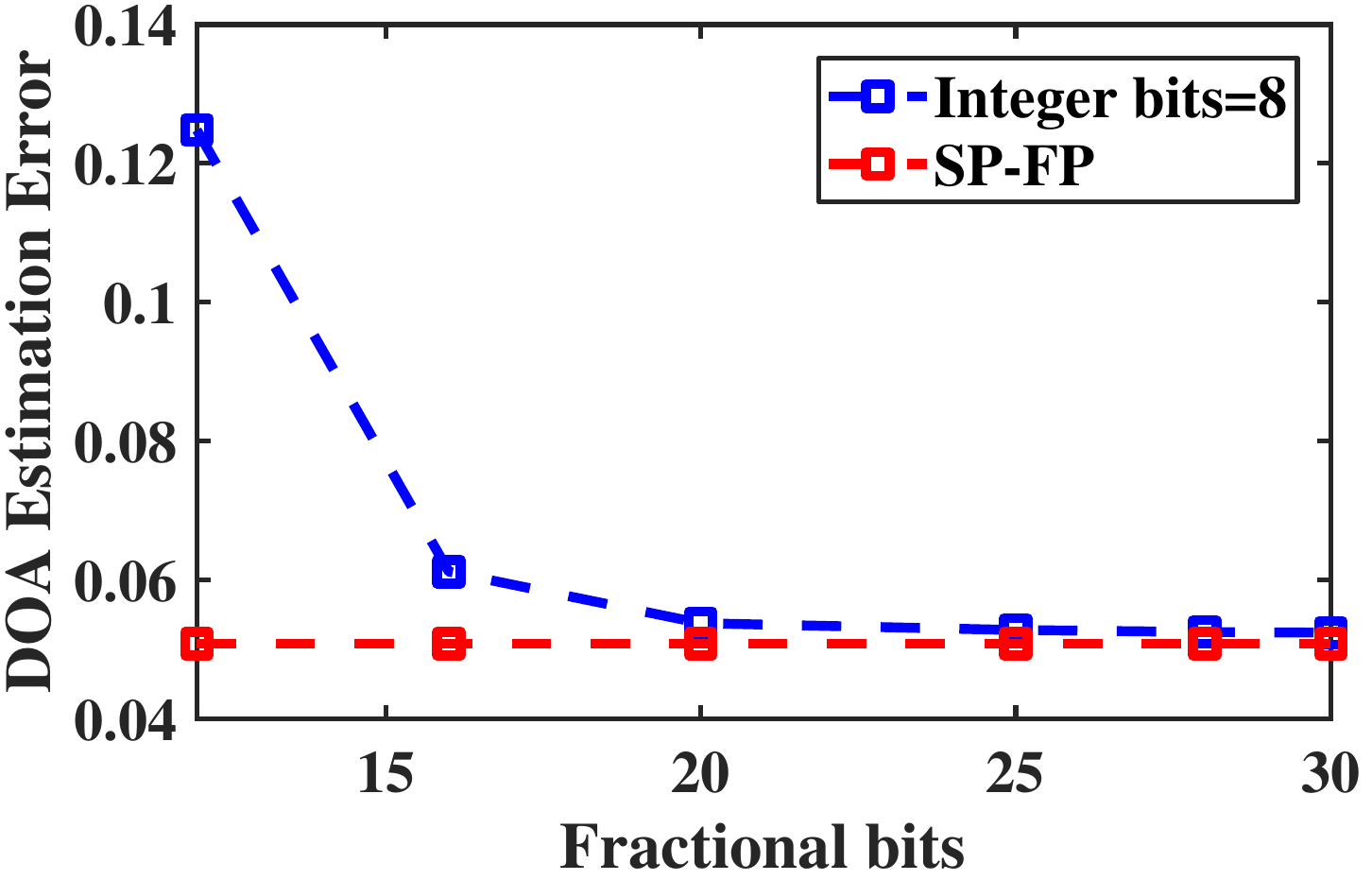}}
        \vspace{-0.1cm}
        \caption{\footnotesize DOA estimation error for varying (a) Integer bits with scaling in ULA (b) Fractional part in ULA (c) Integer bits with scaling in SAA, (d) Fractional part in SAA.}
        \label{fig:ula_integral_part}
\end{figure}

The careful selection of WL of any block of the algorithm realized on FPGA is critical to get the desired trade-off between NDEE and resource utilization. To enable such selection, the proposed algorithms are implemented for different WLs, and corresponding NDEE is analyzed with respect to double-precision FP (DP-FP) i.e. 64 bits FP implementation on software using MATLAB. In Fig.~\ref{fig:ula_integral_part} (a) and (b), the effect on NDEE due to the different number of integer and fractional digits, respectively, is presented for ULA. We have also explored the effect of scaling at the input and outputs of various blocks so as to minimize the number of integer bits. Similar study is repeated for SAA in Fig.~\ref{fig:ula_integral_part} (c) and (d). As expected, NDEE decreases with the increase in the WL. The effect of scaling is significant on reducing the NDEE for a given number of integer bits. Fewer integer bits allow a higher number of fractional bits and hence, smaller NDEE. Interestingly, NDEE does not improve significantly after certain WL which in turn helps to optimize resource and power consumption. Such study can not be done analytically due to complex algorithms and hence, the proposed approach of analyzing NDEE directly on hardware is preferred though time-consuming. Next, the resource utilization and latency of ULA and SAA architectures for three different WLs is compared in Table \ref{tab:resource_utilization}. Here, \{17,7\} represents the fixed point data-type with total of $17$ bits comprising of $7$ integer bits and $10$ fractional bits. Overall resource utilization and power consumption decrease significantly with a decrease in WL. Depending on the application NDEE tolerance, appropriate WL can be selected.

\begin{table}[h]
\vspace{-0.2cm}
    \caption{\footnotesize Complexity Comparison for Different WL}
   \renewcommand{\arraystretch}{1}
         \resizebox{0.5\textwidth}{!}{
    \begin{tabular}{|c|c|c|c|c|c|c|c|}
        \hline
        & \textbf{WL} & \textbf{NDEE} & \textbf{Latency} & \textbf{\{BRAM, DSP, FF, LUT\}} & \textbf{Total \& Dynamic}\\
        & & &  & & \textbf{ Power (in W)} \\
        \hline
        \multirow{3}{*}{\textbf{ULA}}  &SP-FP & 0.065 & 6604 & \{ 34, 130, 23187, 21125 \} & \{ 2.181, 2.023 \}\\
        
        \cline{2-6}
        &24,8 & 0.086 & 6599 & \{ 30, 119, 18700, 19422 \}  & \{ 2.152, 1.995 \} \\ 
        \cline{2-6}
        &17,7 & 0.2 & 6597 & \{ 27, 103, 18003, 17792 \} & \{ 2.073, 1.919 \} \\
        \hline
        \hline
        \multirow{3}{*}{\textbf{SAA}} &SP-FP & 0.050 & 9704 & \{ 98, 176, 31393, 31825 \} & \{ 2.424, 2.254 \}\\
      
        \cline{2-6}
        &24,8 & 0.061 & 9570 & \{ 68, 167, 21815, 28054 \} & \{ 2.246, 2.084 \} \\
          \cline{2-6}
        &17,7 & 0.18 & 9570 & \{ 50, 127, 20836, 23954 \} & \{ 2.237, 2.074 \}\\
        \hline
    \end{tabular}}
    \label{tab:resource_utilization}
    \vspace{-0.2cm}
\end{table}
 

The architectures in Table~\ref{tab:resource_utilization} are designed to minimize the latency. It is possible to serialize the architecture to minimize resource utilization at the cost of latency. For instance, SP-FP architecture for SAA can be designed to reduce the BRAM, DSP, FF, and LUT utilization to 72, 145, 23446, and 22719 from 98, 176, 31393, and 31825, respectively. However, this leads to an increase in latency from 9704 to 24005. The discussion in this paper is limited to architecture exploiting the parallelism in FPGA so as to reduce latency.




Next, we compare the execution time of various configurations of the proposed architecture obtained via HSCD on ZSoC consisting of PS (ARM + NEON) and PL (FPGA). Here, the NEON co-processor exploits single instruction multiple data (SIMD) architecture to accelerate the correlation and matrix operations. As shown in Table~\ref{tab:HSCD_res}, a total of four ULA and five SAA configurations are explored. Here, blocks are numbered as 1) SAP, 2) ACF, 3) EVD, and 4) MSG.
The execution time improves as we accelerate various blocks of the architecture on FPGA. It is possible to achieve up to 54\% improvement in the execution time by adding FPGA on SoC. Without FPGA, the execution time of PS-based configuration can be improved by up to 30\% with the inclusion of NEON co-processor. As expected, FPGA size and power consumption increases as more blocks are realized on FPGA. Depending on the desired resource, latency, power, and cost constraints, an appropriate configuration can be chosen from Table~\ref{tab:HSCD_res}.

        \begin{table}[!h]
        \centering
        \vspace{-0.1cm}
         \caption{\footnotesize Comparison for Various HSCD Configurations on Zynq SoC}
         \renewcommand{\arraystretch}{1}
         \resizebox{0.5\textwidth}{!}{
        \begin{tabular}{|c|c|c|c|c|}
        \hline
         \textbf{Blocks} & \textbf{ZSoC} & \textbf{ZSoC+} & &\textbf{Power (Total} \\
         \textbf{in PL} & \textbf{} & \textbf{NEON} &\textbf{\{ BRAM, DSP, FF, LUT \}}&\textbf{\& Dynamic)}\\
         & \textbf{(in us)} & \textbf{(in us)} &&(in W)\\
        \hline
           
       \textbf{NA}& 143.11 & 90 & NA &NA \\
       \cline{1-5}
       \textbf{2} & 135.81 & 92.24 & \{  8,  25,  1740,  1717 \} &\{ 1.74, 1.60 \} \\
       \cline{1-5}
       \textbf{2-3} & 130.33 & 94.57 & \{ 20, 98, 17081, 16097 \} & \{ 2.05, 1.90 \}  \\
         \cline{1-5}
     \textbf{2-4} & 74.70 & NA & \{ 34, 130, 23187, 21125 \} & \{ 2.18, 2.02 \}\\
      \hline
      \hline
      \textbf{NA}& 284.77& 184.49&NA&NA\\
       \cline{1-5}
      \textbf{1} &283.17 & 193.09 & \{ 22, 24, 2239, 2990 \} & \{ 1.77, 1.63 \}\\
    \cline{1-5}
    \textbf{1-2}& 278.80 & 190.71 & \{ 42, 54, 6273, 6867 \} &\{ 2.05, 1.90 \}\\
    \cline{1-5}
    \textbf{1-3} & 259.06 & 182.41 & \{ 68, 128, 21511, 21624 \} & \{ 2.17, 2.01 \} \\
    \cline{1-5}
    \textbf{1-4} & 131.44 &NA& \{ 98, 176, 31393, 31825 \} & \{ 2.42, 2.25 \} \\
    \hline
    \end{tabular}}
    \label{tab:HSCD_res}
    \vspace{-0.1cm}
    \end{table}

Next, we design the reconfigurable architecture which can dynamically switch between algorithms depending on the value of $M$. Specifically, $V_n$ extraction from the output of QR decomposition block in EVD and MSG are reconfigured depending on the given $M$. In Table \ref{tab:dpr_ula_sparse}, we consider three cases: 1) ULA with $M=\{1,2\}$, 2) SAA with $M=\{2,4,5\}$, and 3) SAA with $M=\{1,2,3, 4,5\}$. In a non-DPR-based implementation, instead of realizing the separate $V_n$ extraction and MSG block for each $M$, we have exploited the redundancy between different $M$ so as to reduce resource and power utilization. Still, DPR-based architecture offers lower resource utilization and significant power savings over non-DPR-based architecture. The savings increase significantly with an increase in $|M|$. Furthermore, DPR-based architecture is faster with the minimum clock period of $10ns$ compared to $16ns$ in non-DPR-based architecture.

    \begin{table}[!h]
        \vspace{-0.1cm}
        \caption{\footnotesize Comparison of DPR and Non-DPR based Architectures.}
        \renewcommand{\arraystretch}{1}
         \resizebox{0.5\textwidth}{!}{
        \begin{tabular}{|c|c|c|c|c|c|c|c|}
            \hline
             &  & \textbf{Total} & \textbf{Dynamic} & &  &&  \\
           \textbf{Case} &\textbf{Approach} & \textbf{Power} & \textbf{Power} & \textbf{BRAM}&\textbf{DSP}& \textbf{LUTs} & \textbf{FF}\\
            & & (in W) & (in W) & & & & \\
            \hline
            \multirow{3}{*}{ULA} & DPR & \textbf{2.12} & \textbf{1.96} & 20 & \textbf{124} & \textbf{20950} & \textbf{26396}
            \\
            & & \textbf{(-9.01\%)} & \textbf{(-9.26\%)} & & \textbf{(-39\%)} & \textbf{(-30\%)} & \textbf{(-29.7\%)}
            \\
            \cline{2-8} 
            $M$=1-2 & No-DPR & 2.33 & 2.16 & 19 & 203 & 29930 & 37551 \\
            
            \hline
            \multirow{2}{*}{SAA} & DPR & 2.40 & 2.23 & 64 & 179 & \textbf{31724} & \textbf{35267}
            \\
            & & & & & & \textbf{(-13\%)} & \textbf{(-8.45\%)} \\
            \cline{2-8} 
            $M=$2,4,5 & No-DPR & 2.39 & 2.22 & 58.5 & 176 & 36455 & 38522 \\
            \hline
            \multirow{2}{*}{SAA} & DPR & \textbf{2.42} & \textbf{2.25} & 62 & 179 & \textbf{22124} & \textbf{34941}
            \\
            & & \textbf{-1.62\%} & \textbf{-1.74\%} & & & \textbf{(-52.9\%)} & \textbf{(-22.5\%)} \\
            \cline{2-8} 
            $M$=1-5 & No-DPR & 2.46 & 2.29 & 62 & 178 & 46986 & 45068 \\
            \hline
        \end{tabular}}
        \vspace{-0.1cm}
        \label{tab:dpr_ula_sparse}
    \end{table}

.

\vspace{-0.2cm}

\section{Conclusions and Future Works}
\label{Sec:conclude}
In this paper, we proposed reconfigurable architectures for spatial sensing in wideband radio front-end (WRFE) comprising of Sub-Nyquist Sampling (SNS) and sparse antenna array (SAA). We have demonstrated the functional correctness of the architecture for different signal-to-noise ratios (SNR), samples, and word-length (WL). We analyzed the effect of WL and various configurations obtained via hardware-software co-design on resource utilization, power, and execution time. The advantages of the reconfigurable architecture were also demonstrated. Future work includes the extension of spatial sensing for unknown carrier frequency, exploring a computationally efficient alternative to the MUSIC algorithm, and integration with RF front-end for demonstration in the real-radio environment.

\newpage

    

\section*{Supplementary material}

# Tutorial: Reconfigurable DOA Estimation using MUSIC Algorithm via HLS

## Introduction

In this lab, you will use Vivado HLS, Vivado IP Integrator and Software Development Kit to create a reconfigurable peripheral using ARM Cortex-A9 processor system on Zynq. You will use Vivado HLS to generate the user-defined IPs, Vivado IPI to create a top-level design which includes the Zynq processor system as a sub-module. During the PR flow, you will define one Reconfigurable Partition which will have two modes. You will create multiple Configurations and run the Partial Reconfiguration implementation flow to generate full and partial bitstreams. You will use Zedboard to verify the design in hardware using a SD card to initially configure the FPGA, and then partially reconfigure the device using the PCAP under user software control.

**GitHub Link:** https://github.com/Somya-Sharma/SpatialSensing_MUSIC.git

## Design Description

The purpose of this lab exercise is to implement a design that can dynamically reconfigure number of active DOA in Spatial sensing using PCAP resource and PS sub-system. This is done for two types of array arrangements, viz. Uniform Linear Array (ULA) and Sparse Array Arrangement (SAA). The proposed reconfigurable architecture for on ZSoC is shown in Fig. **??**. The data stored by the ARM Core 0 of the PS in the DDR memory. The AXI Direct Memory Access (DMA) reads the data from DDR memory via memory-mapped AXI Accelerator Coherency Port (ACP), forwards it to spatial sensing block via AXI stream interface and stores the processed data back to the DDR memory. The PS displays the calculated DoA using UART terminal which is used to verify the functionality of the design. The ULA spatial sensing architecture is obtained by removing the SAP block in Fig. **??**. The architecture is made reconfigurable in $M$ i.e. number of active transmissions, by DPR based on-the-fly configuration of Extract $\mathbf{V}_n$ and Music Spectra Generation blocks via processor configuration access port (PCAP).

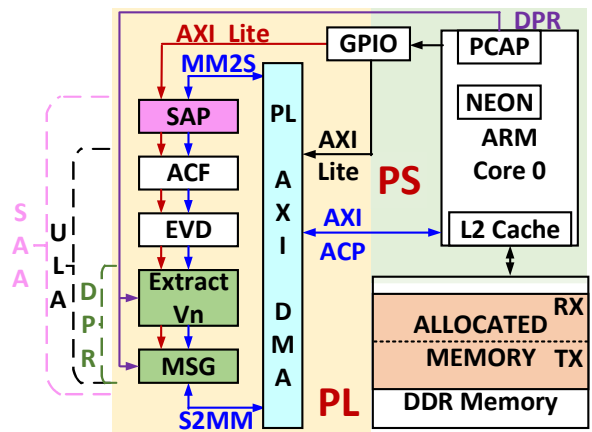

Figure 1: Proposed reconfigurable architecture for spatial sensing on Zynq SoC.

## General Flow for this Lab

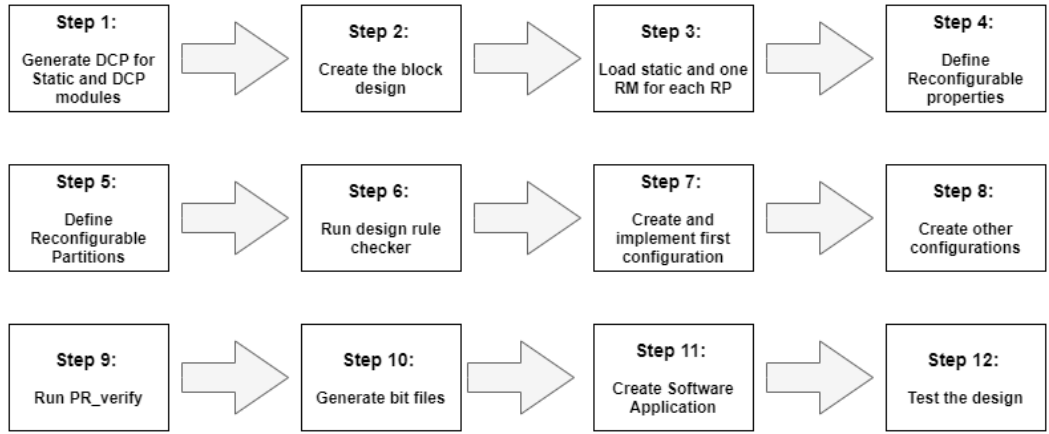

## Procedure

This lab is separated into steps that consist of general overview statements that provide information on the detailed instructions that follow. Follow these detailed instructions to progress through the lab.

## Generate DCPs for Static and RM Modules             Step 1

**1-1. Start Vivado and execute the provided Tcl script to create the design check point for the static design having one RP.**

1-1-1. Open Vivado by selecting **Start > All Programs > Xilinx Design Tools > Vivado 2019.1.**

1-1-2. In the Tcl Shell window enter the following command to change to the lab directory and hit **Enter**.

<span style="color:red">For SAA, migrate to `C:/DPR_MUSIC/Tutorial/SAA`</span>
<span style="color:blue">For ULA, migrate to `C:/DPR_MUSIC/Tutorial/ULA`</span>

1-1-3. Generate the PS design using given TCl script.

```
source generate_bd.tcl
```

This script will create a new project *DPR_proj* and the block design *bd_dpr*, instantiate ZYNQ PS with SD 0 and UART 1 interfaces enabled. It will also enable the GP0 interface along with FCLK0 and RESET0_N ports. The provided IP for static and reconfigurable region will also be instantiated. It will then create a top-level wrapper file called *bd_dpr_1_wrapper.v* which instantiates the *bd_dpr.bd* (the block design).

1-1-4. Select the **Address Editor** tab. Expand the design heirarchy. Expand **Unmapped Slaves**, if any, and right-click and select **Auto-Assign Address**. Similarly, Expand **Excluded segment**, if any, and right-click and select **Include segment**.

1-1-5. Select **Tools > Validate Design.**

1-1-6. Select **File > Save block design.**

---

**1-2. The current directory contains a folder named Netlists where all the design checkpoints (dcp) will be maintained. It further contains sub folders Static, Config1, Config2, and blank. Synthesize the design to generate the dcp for the static logic of the design.**

1-2-1. Click **Run Synthesis** under the Synthesis group in the Flow Navigator to run the synthesis process. Wait for the synthesis to complete. When done click **Cancel**.

1-2-2. Using the windows explorer, copy the *bd_dpr_1_wrapper.dcp* file from *./bd_dpr/bd_dpr.runs/synth_1* into the *./Netlists/Static* directory.

1-2-3. Copy design checkpoints for the *processing_system7_0* instances to the *./Netlists/Static*.

## Define the Reconfigurable Partition region                                Step 2

**2-1. Next you must floorplan the RP region. Depending on the type and amount of resources used by the RMs for the given RP, the RP region must be appropriately defined so it can accommodate any RM variant. For our design, minimum of two clock regions are required for SAA and only one clock region for ULA.**

2-1-1. You execute the following TCl script to assign a pblock to the reconfigurable instance and set the properties of hence created pblock.

```
source create_pblock.tcl
```

## Load Static and one RM Module                                             Step 3

**Since all required netlist files (dcp) for the design are already given in the Synth folder, you will use Vivado to floorplan the design, define Reconfigurable Partitions, add Reconfigurable Modules, run the implementation tools, and generate the full and partial bitstreams.**

**3-1. In this step you will load one of the RM designs for the RP.**

3-1-1. Since the project is already open, we do not need to load the static design checkpoint. Load the first configuration using the **read_checkpoint** command to the reconfigurable instance.

For SAA, `read_checkpoint -cell [get_cells bd_dpr_i/SAA_Recon_0/inst] ./Netlists/Config1/config1.dcp`

For ULA, `read_checkpoint -cell [get_cells bd_dpr_i/ULA_Recon_0/inst] ./Netlists/Config1/config1.dcp`

---

### 3-2. In this design you have only one Reconfigurable Partition. Define the reconfigurable properties to the loaded RM.

3-2-1. Define loaded RM (submodule) as partially reconfigurable by setting the HD.RECONFIGURABLE property using the following command.

For SAA, `set_property HD.RECONFIGURABLE 1 [get_cells bd_dpr_i/SAA_Recon_0/inst]`

For ULA, `set_property HD.RECONFIGURABLE 1 [get_cells bd_dpr_i/ULA_Recon_0/inst]`

This is the point at which the Partial Reconfiguration license is checked.

## Create and implement first configuration      Step 4

### 4-1. Create and implement the first Configuration.

4-1-1. Execute the following command to implement the first configuration. For SAA it is $M = 2$ while for ULA, it is $M = 1$.

```
source config1_route.tcl
```

The script does the following tasks.

- Write the pre-routing design checkpoint for this configuration.

  `write_checkpoint -force ./Netlists/Config1/static1.dcp`

- Optimize, place and route the design by executing the following commands.

  `opt_design`

  `place_design`

  `route_design`

- Finally write the post-routing design checkpoint.

  `write_checkpoint -force ./Netlists/Config2/static2_routed.dcp`

### 4-2. After the first configuration is created, the static logic implementation will be reused for the rest of the configurations. So it should be saved. But before you save it, the loaded RM should be removed.

4

4-2-1. Execute the following script to update the design with reconfigurable instance as black box.

```
source lock_static.tcl
```

The script will do the following tasks.

- Clear out the existing RMs executing the following commands.

  For SAA, `,update_design -cell [get_cells bd_dpr_i/SAA_Recon_0/inst] -black_box`

  For ULA, `,update_design -cell [get_cells bd_dpr_i/ULA_Recon_0/inst] -black_box`

  Issuing this command will result in design changes including, the number of Fully Routed nets (green) decreased, the number of Partially Routed nets (yellow) has increased, and RPs may appear in the Netlist view as empty.

- Lock down all placement and routing by executing the following command.

  ```
  lock_design -level routing
  ```

  Because no cell was identified in the lock_design command, the entire design in memory (currently consisting of the static design with black boxes) is affected.

- Write out the remaining static-only checkpoint by executing the following command.

  ```
  write_checkpoint -force ./Netlists/Static/static_routed_design.dcp
  ```

  This static-only checkpoint would be used for any future configuration, but here, you simply keep this design open in memory.

# Create other configurations                                         Step 5

### 5-1. Read next RM dcp, create and implement the second configuration.

5-1-1. Execute the following command to create and implement the second configuration. For SAA it is $M = 4$ while for ULA, it is $M = 2$.

For SAA, `read_checkpoint -cell [get_cells bd_dpr_i/SAA_Recon_0/inst] ./Netlists/Config2/config2.dcp`

For ULA, `read_checkpoint -cell [get_cells bd_dpr_i/ULA_Recon_0/inst] ./Netlists/Config2/config2.dcp`

5-1-2. Execute the following TCl script to create the routed design checkpoint for current configuration.

```
source config2_route.tcl
```

The script does following tasks.

- Write the pre-routing design checkpoint for this configuration.

    ```
    write_checkpoint -force ./Netlists/Config2/static2.dcp
    ```

- Optimize, place and route the design by executing the following commands.

    ```
    opt_design

    place_design

    route_design
    ```

- Finally write the post-routing design checkpoint.

    ```
    write_checkpoint -force ./Netlists/Config2/static2_routed.dcp
    ```

# Generate bitfiles for created configurations — Step 6

## 6-1. After all the Configurations have been validated by PR_Verify, full and partial bit files must be generated for the entire project.

6-1-1. The design for the second configuration is already opened. So, there is no need to lead any design checkpoint and you you can directly generate the bitstream for this conifguration.

```
source config2_bitstream.tcl
```

The script does the following tasks.

- Generate the bitstream.

    ```
    write_bitstream -force ./Bitstreams/Config2/config2.bit
    ```

    This command will generate full and partial bitstream for the current configuration.

- Generate the corresponding bin file (of the partial bit file) which will be used to program the FPGA via the SD card.

    ```
    write_cfgmem -force -format BIN -interface SMAPx32 -disablebitswap
    -loadbit "up 0x0 ./Bitstreams/Config2/config2_pblock_inst_partial.bit"
    "./Bin_files/config2.bin"
    ```

6-1-1. Open the design checkpoint for the the first created configuration by executing the below command.

```
open_checkpoint ./Netlists/Config1/static1_routed.dcp
```

6-1-2. Execute the following TCL script in the new window that opens.

```
source config1_bitstream.tcl
```

The script does the following tasks.

- Generate the bitstream.

    ```
    write_bitstream -force ./Bitstreams/Config1/config1.bit
    ```

- Generate the corresponding bin file which will be used to program the FPGA via the SD card.

    ```
    write_cfgmem -force -format BIN -interface SMAPx32 -disablebitswap
    -loadbit "up 0x0 ./Bitstreams/Config1/config1_pblock_inst_partial.bit"
    "./Bin_files/config1.bin"
    ```

- Close the project.

    ```
    close_project
    ```

## Create blank configuration                                                         Step 7

**7-1. Create the blanking configuration.**

7-1-1. Open the design checkpoint for the the blank configuration (which is routed static region) executing the below command.

```
open_checkpoint ./Netlists/Static/static_routed.dcp
```

7-1-2. Execute the following Tcl script to create the blank configuration i.e. no module is loaded to the reconfigurable partition.

```
source blank_bitstream.tcl.
```

The script does the following tasks.

- For creating the blanking configuration, use the update_design -buffer_ports command to insert LUTs tied to constants to ensure the outputs of the reconfigurable partition are not left floating.

    For SAA, `update_design -buffer_ports -cell bd_dpr_i/SAA_Recon_0/inst`
    For ULA, `update_design -buffer_ports -cell bd_dpr_i/ULA_Recon_0/inst`

- Now place and route the design. There is no need to optimize the design.

  `place_design`

  `route_design`

  The base (or blanking) configuration bitstream, when we generate in the next section, will have no logic for either reconfigurable partition, simply outputs driven by ground. Outputs can be tied to VCC if desired, using the HD.PARTPIN_TIEOFF property.

- Save the checkpoint in the BLANK directory.

  `write_checkpoint -force ./Netlists/Blank/blank_routed.dcp`

# Export the hardware and open SDK                                    Step 8

### 8-1. Export the architecture as a hardware definition file.

8-1-1. Execute the following TCl script for this step.

  `source exportHW.tcl`

  This script does the following tasks

  - Create the .sdk folder to where the user application will be created.

    For SAA, `file mkdir C:/DPR_MUSIC/Tutorial/SAA/DPR_proj/DPR_proj.sdk`
    For ULA, `file mkdir C:/DPR_MUSIC/Tutorial/ULA/DPR_proj/DPR_proj.sdk`

  - Write the hardware definition file.

    For SAA, `write_hwdef -force -file C:/DPR_MUSIC/Tutorial/SAA/DPR_proj/DPR_proj.sdk/bd_dpr_wrapper.hdf`
    For ULA, `write_hwdef -force -file C:/DPR_MUSIC/Tutorial/ULA/DPR_proj/DPR_proj.sdk/bd_dpr_wrapper.hdf`

  You can also go to **File > Export Hardware**, leave the **Include Bitstream** option **unchecked**, click **OK** for this step.

8-1-2. Select **File > Launch SDK**

8-1-3. Click OK to launch SDK. The SDK program will open. Close the Welcome tab if it opens

# Create the software application                                     Step 9

### 9-1. Create a Board Support Package enabling generic FAT file system library.

9-1-1. In SDK, select **File > New > Board Support Package**.

9-1-2. Click **Finish** with the default settings (with standalone operating system). This will open the Software Platform Settings form showing the OS and libraries selections.

9-1-3. Select **xilffs** as the FAT file support is necessary to read the partial bit files from the SD card.

9-1-4. Click OK to accept the settings and create the BSP.

---

### 9-2. Create an application.

9-2-1. Select **File > New > Application project**.

9-2-2. Enter TestApp as the Project Name, and for Board Support Package, choose Use Existing (standalone_bsp_0 should be the only option). If this option does not appear, make sure that the hardware specification is set to the hdf created above in step 8-1-1.

9-2-3. Click **Next**, and select *Empty Application* and click **Finish**.

9-2-4. Expand the **TestApp** entry in the project view, right-click the *src* folder, and select **Import**.

9-2-5. Expand **General category** and double-click on **File System.**.

9-2-6. Browse to C:/DPR_MUSIC/Tutorial/Sources/<SAAorULA> or and click **OK**.

9-2-7. Select *TestApp.c, lib_music.h, lib_music.c, pcap.h, pcap.c, sdCard.h, sdCard.c, platform.h, platform.c, platform_config.h* and **click Finish** to add the file to the project.

9-2-8. Right-click on **TestApp** and select **C/C++ Building Settings**.

---

### 9-3. Create a zynq_fsbl application.

9-3-1. Select **File > New > Application project**.

9-3-2. Enter **zynq_fsbl** as the Project Name, and for Board Support Package, choose **Create New**.

9-3-3. Click **Next**, select **Zynq FSBL**, and click **Finish**.
This will create the first stage bootloader application called **zynq_fsbl.elf**

---

### 9-4. Create a Zynq boot image.

9-4-1. Select **Xilinx Tools > Create Boot Image**.

9-4-2. Click the Browse button of the Output BIF file path field, browse to *C:/DPR_MUSIC/ Tutorial/<SAAorULA>* and then click **Save** with the output as the default filename.

9-4-3. Click on the **Add** button of the Boot image partitions, click the Browse button in the Add Partition form, browse to *C:/DPR_MUSIC/Tutorial/<SAAorULA>/DPR_proj/DPR_proj.sdk/zynq_fsbl/ Debug* select **zynq_fsbl.elf** and click **Open**.

9-4-4. Click **OK**. Click again on the **Add** button of the Boot Image partitions, click the Browse button in the Add Partition form, browse to **C:/DPR_MUSIC/<SAAorULA>/Bitstreams** directory, select BLANK.bit and click **Open**.

9-4-5. Click **OK**.

9-4-6. Click again on the **Add** button of the Boot Image partitions, click the Browse button in the Add Partition form, browse to **C:/DPR_MUSIC/Tutorial/<SAAorULA>/DPR_proj/ DPR_proj.sdk/TestApp/Debug**, select **TestApp.elf** and click **Open**.

9-4-7. Click **OK**.

9-4-8. Make sure that the output path is **C:/DPR_MUSIC/Tutorial/<SAAorULA>/DPR_proj/** and the filename is **BOOT.bin**. Click **Create Image**.

9-4-9. Close the SDK program by selecting **File > Exit**.

# Test the design                                                        Step 10

**10-1. Connect the board with micro-USB cable connected to the UART. Place the board in the SD boot mode. Copy the generated BOOT.bin and the partial bit files on the SD card, and place the SD card in the board. Power On the board.**

10-1-1. Make sure that a micro-usb cable is connected to the UART port.

10-1-2. Make sure that the board is set to boot in SD card boot mode. For this, check that the jumper connections are set to program from the SD card.

10-1-3. Using the Windows Explorer, copy the **BOOT.bin** and other partial binaries on to a SD Card.

10-1-4. Place the SD Card in the board and power ON the board.

**10-2. Start a terminal emulator program such as TeraTerm or HyperTerminal. Select an appropriate COM port (you can find the correct COM number using the Control Panel). Set the COM port for 115200 baud rate communication.**

10-2-1. Start a terminal emulator program such as **TeraTerm** or **Terminal**.

10-2-2. Select the appropriate COM port (you can find the correct COM number using the Control Panel).

10-2-3. Set the COM port for **115200** baud rate communication.

10-2-4. Press BTN7 to display a menu.

10-2-5. Choose either 2 or 4 for SAA and 1 or 2 for ULA.

10-2-6. Below in Fig. 2 is an example user test to show how the terminal window will appear after various reconfigurations.

(a) SAA

(b) ULA

Figure 2: DPR Result for ULA and SAA



\end{document}